\documentclass[aps,prd,amsmath,superscriptaddress,twocolumn]{revtex4}

\usepackage{graphicx}
\usepackage{multirow}

\def\gae{\lower 2pt \hbox{
 $\, \buildrel {\scriptstyle >}\over {\scriptstyle \sim}\,$}}
\def\lae{\lower 2pt \hbox{
 $\, \buildrel {\scriptstyle <}\over {\scriptstyle \sim}\,$}}
\newcommand{\erf}{\mathop{\rm erf}}

\begin{document}


\preprint{MCTP-04-47}


\title{Can WIMP Spin Dependent Couplings explain DAMA data,
in light of Null Results from Other Experiments?}

\author{Chris Savage}
\affiliation{
 Michigan Center for Theoretical Physics,
 Physics Department,
 University of Michigan,
 Ann Arbor, MI 48109}

\author{Paolo Gondolo}
\affiliation{
 Physics Department,
 University of Utah,
 Salt Lake City, UT 84112}

\author{Katherine Freese}
\affiliation{
 Michigan Center for Theoretical Physics,
 Physics Department,
 University of Michigan,
 Ann Arbor, MI 48109}

\date{\today}


\begin{abstract}
  
  We examine whether the annual modulation found by the DAMA dark
  matter experiment can be explained by Weakly Interacting Massive
  Particles (WIMPs), in light of new null results from other
  experiments.  CDMS II has already ruled out most WIMP-nucleus
  spin-independent couplings as an explanation for DAMA data. Hence we
  here focus on spin-dependent (axial vector; SD) couplings of WIMPs
  to nuclei.  We expand upon previous work by (i) considering the
  general case of coupling to both protons and neutrons and (ii)
  incorporating bounds from all existing experiments.  We note the
  surprising fact that CMDS II places one of the strongest bounds on
  the WIMP-neutron cross-section, and show that SD WIMP-neutron
  scattering alone is excluded.  We also show that SD WIMP-proton
  scattering alone is allowed only for WIMP masses in the 5-13 GeV
  range.  For the general case of coupling to both protons and
  neutrons, we find that, for WIMP masses above 13 GeV and
  below 5 GeV, there is no region of parameter space that is
  compatible with DAMA and all other experiments.  In the range
  (5-13) GeV, we find acceptable regions of parameter space,
  including ones in which the WIMP-neutron coupling is comparable to
  the WIMP-proton coupling.

\end{abstract}

\maketitle


\section{\label{sec:Intro} Introduction}

The dark halo of our Galaxy may consist of WIMPs (Weakly Interacting
Massive Particles).  Numerous collaborations worldwide have been
searching for these particles. Direct detection experiments attempt to
observe the nuclear recoil caused by these dark matter particles
interacting with nuclei in the detectors.  Indirect detection
experiments search for the signatures of WIMP annihilation in the Sun,
the Earth, or the Galactic Halo.  It is well known that the count rate
in WIMP direct detection experiments will experience an annual
modulation \cite{Drukier:1986tm, Freese:1987wu} as a result of the
motion of the Earth around the Sun: the relative velocity of the
detector with respect to the WIMPs depends on the time of year.

The discovery of an annual modulation by the DAMA/NaI experiment
\cite{Bernabei:2003za} (hereafter, ``DAMA'') is the only positive
signal seen in any dark matter search.  However, recent null results
from other experiments, particularly EDELWEISS and CDMS II, severely
bound the parameter space of WIMPs that could possibly explain the
DAMA data.  In fact, if the interactions between WIMPs and nuclei are
spin-independent (SI), then CDMS II has ruled out DAMA (with some
exceptions \cite{Gelmini:2004gm}).  In this paper, we will consider the
alternative of spin-dependent (SD) WIMP cross-sections.
We expand upon previous work by (i) considering the general case of
coupling to both protons and neutrons and (ii) incorporating bounds
from all existing experiments, including, in particular, the null
results recently released by CDMS II and Super-Kamiokande (Super-K),
which were not available to previous authors.  Previously, Ullio
\textit{et al.}\ studied the cases of interactions with protons only or
neutrons only \cite{Ullio:2000bv} and made a comparison with experiments
that had been performed at that time; they argued that some cases of
mixed interactions would be ruled out as well.  In 2003 Kurylov and
Kamionkowski performed a general analysis for point-like WIMPs of
arbitrary spin to find what WIMPs were compatible with DAMA and the
null experiments at the time \cite{Kurylov:2003ra}.  Giuliani looked at
the more general case of mixed interactions \cite{Giuliani:2004uk}, but
only applied the analysis to DAMA for a limited choice of WIMP mass.  We
here perform a general analysis with coupling to both protons and
neutrons to see if any parameter space remains that could explain
simultaneously the DAMA and CDMS data.  To be complete, we will also
incorporate bounds from other experiments, including Super-K
(which measures indirect detection in the Sun).  Notice, however, that
these indirect detection limits do not apply if WIMPs do not effectively
annihilate in the Sun, as for example if the WIMP is not its own
anti-particle and there is a sufficiently strong cosmic asymmetry in the
number of WIMPs and anti-WIMPs.

We find those regions of WIMP parameter space that survive all
existing experiments for the case of spin-dependent interactions.
Although the current CDMS data are more stringent than those studied
in previous work, our final bounds are somewhat less restrictive for
WIMP-proton coupling than those quoted by Ullio \textit{et al.}\ 
because the latest results released by Super-K are less restrictive
than those approximated by Ullio \textit{et al.}\ and because we are
reluctant to extend Super-K data beyond those published by the
experimentalists (as elaborated in the Discussion section at
the end of the paper).

We also note that, although CDMS data have typically been ignored in
the spin-dependent sector due to the small natural abundance of spin
odd isotopes in the detector, in fact CDMS II places one of the
strongest bounds on the WIMP-neutron cross-section (\textit{n.b.} even
CDMS I placed interesting bounds particularly at low masses).

\section{\label{sec:Detection} WIMP Detection}

The basic idea underlying WIMP direct detection is straightforward: the
experiments seek to measure the energy deposited when a WIMP interacts
with a nucleus in the detector \cite{Goodman:1984dc}.

If a WIMP of mass $m$ scatters elastically from a nucleus of mass $M$,
it will deposit a recoil energy $E = (\mu^2v^2/M)(1-\cos\theta)$,
where $\mu \equiv m M/ (m + M)$ is the reduced mass, $v$ is the speed
of the WIMP relative to the nucleus, and $\theta$ is the scattering
angle in the center of mass frame.  The differential recoil rate per
unit detector mass for a WIMP mass $m$, typically given in units of
counts/kg/day/keV, can be written as:
\begin{equation} \label{eqn:dRdE}
 \frac{dR}{dE} = \frac{\rho}{2 m \mu^2}\, \sigma(q)\, \eta(E,t)
\end{equation}
where $\rho = 0.3$ GeV/cm$^3$ is the standard local halo WIMP density,
$q = \sqrt{2 M E}$ is the nucleus recoil momentum, $\sigma(q)$ is the
WIMP-nucleus cross-section, and  information about the WIMP velocity
distribution is encoded into the mean inverse speed $\eta(E,t)$,
\begin{equation} \label{eq:eta}  
  \eta(E,t) = 
  \int_{u>v_{\rm min}} \frac{f_{\rm d}({\bf u},t)}{u} \, d^3u .
\end{equation}
Here 
\begin{equation} \label{eq:vmin}
  v_{\rm min} = \sqrt{\frac{M E}{2\mu^2}}
\end{equation}
represents the minimum WIMP velocity that can result in a recoil energy
$E$ and $f_{\rm d}({\bf u},t)$ is the (usually time-dependent)
distribution of WIMP velocities ${\bf u}$ relative to the detector.

For WIMPs in the Milky Way halo, the most frequently employed WIMP
velocity distribution is that of a simple isothermal sphere
\cite{Freese:1987wu}.  In such a model, the Galactic WIMP speeds with
respect to the halo obey a Maxwellian distribution with a velocity
dispersion $\sigma_h$ truncated at the escape velocity $v_{\rm esc}$,
\begin{equation}
  f_h(\mathbf{v}) =
    \begin{cases}
      \frac{1}{N_{\textrm{esc}} \pi^{3/2} \overline{v}_0^3} 
        \, e^{-\mathbf{v}^2\!/\overline{v}_0^2} , 
        & \textrm{for} \,\, |\mathbf{v}| < v_{\textrm{esc}}  \\
      0 , & \textrm{otherwise}.
    \end{cases}
\end{equation}
Here 
\begin{equation}
  \overline{v}_0=\sqrt{2/3} \, \sigma_h
\end{equation}
and 
\begin{equation}
  N_{\rm esc} = \erf(z) - 2 z \exp(-z^2) / \pi^{1/2} ,   
\end{equation}
with $z = v_{\rm esc}/\overline{v}_0$, is a normalization factor.  For
the sake of illustration, we take $\sigma_h = 270$ km/s and $v_{\rm
 esc} = 650$ km/s.  The results for the mean inverse speed $\eta(E,t)$
for the case of the isothermal sphere have previously been calculated
and can be found e.g.\ in \cite{Drukier:1986tm,Gelmini:2000dm}.
 
It is well known that the count rate in WIMP detectors will experience
an annual modulation \cite{Drukier:1986tm, Freese:1987wu} as a result
of the motion of the Earth around the Sun: the relative velocity of the
detector with respect to the WIMPs depends on the time of year.  The
disk of the galaxy rotates through the relatively stationary halo,
giving the Sun a velocity $v_{\odot} = 232$ km/s relative to the halo.
In addition, the Earth travels about the Sun at $V_{\oplus} = 30$ km/s
(relative to the Sun), with the Sun's motion in the Galaxy being at 
$60\deg$ from the Earth's orbital plane (${\bf v}_{\oplus} \equiv
{\bf v}_{\odot} + {\bf V}_{\oplus}$).

We will use the isothermal halo model in our calculations, but its
validity is by no means guaranteed -- the actual halo may be squashed or
anisotropic \cite{bib:halomodel}, may contain streams from smaller
galaxies in the process of  being absorbed by the Milky Way
\cite{bib:sagstream} or may, in fact, not yet be thermalized.

An alternative way to search for WIMP dark matter of relevance to this
paper is via indirect detection of WIMP annihilation in the Sun.  When
WIMPs pass through a large celestial body, such as the Sun or the Earth
\cite{indirectdet:solar,indirectdet:earth}, interactions can lead to
gravitational capture if enough energy is lost in the collision to fall
below the escape velocity.  Captured WIMPs soon fall to the body's core
and eventually annihilate with other WIMPs.  These annihilations lead
to high-energy neutrinos that can be observed by Earth-based detectors
such as Super-Kamiokande \cite{Desai:2004pq}, Baksan \cite{bib:baksan},
and AMANDA \cite{Ahrens:2002eb}, and the IceCube \cite{Ahrens:2002dv}
and ANTARES \cite{Blanc:2003na} projects.  The annihilation rate
depends on the capture rate of WIMPs, which is in turn determined by
the WIMP scattering cross section off nuclei in the celestial body.
While the Earth is predominantly composed of spinless nuclei, the Sun
is mostly made of hydrogen, which has spin. Thus the spin-dependent
cross section of WIMPs off nucleons can be probed by measuring the
annihilation signals from WIMP annihilation in the Sun.  Other indirect
detection methods search for WIMPs that annihilate in the Galactic Halo
or near the Galactic Center where they produce neutrinos, positrons, or
antiprotons that may be seen in detectors on the Earth
\cite{indirectdet:galactichalo,indirectdet:galacticcenter}.

\section{\label{sec:SICS} Spin-Independent Cross-Section}

The cross-section for spin-independent WIMP interactions is given by
\begin{equation} \label{eqn:SICS}
 \sigma_{SI} = \sigma_0 \, F^2(q)
\end{equation}
where $\sigma_0$ is the zero-momentum WIMP-nuclear cross-section and
$F(q)$ is the nuclear form factor, normalized to $F(0) = 1$; a
description of these form factors may be found in \cite{SmithLewin} and
\cite{Jungman:1995df}.  For purely scalar interactions,
\begin{equation} \label{eq:scalar}
 \sigma_{0,\rm scalar} = \frac{4 \mu^2}{\pi} [Zf_p + (A-Z)f_n]^2 \, .
\end{equation}
Here $Z$ is the number of protons, $A-Z$ is the number of neutrons,
and $f_p$ and $f_n$ are the WIMP couplings to the proton and nucleon,
respectively.  In most instances, $f_n \sim f_p$; the WIMP-nucleus
cross-section can then be given in terms of the WIMP-proton
cross-section as a result of Eqn.~(\ref{eq:scalar}):
\begin{equation} \label{eqn:SICSproton}
 \sigma_0 = \sigma_p \left( \frac{\mu}{\mu_p} \right)^2 A^2
\end{equation}
where the $\mu_p$ is the proton-WIMP reduced mass, and $A$ is
the atomic mass of the target nucleus.

In this model, for a given WIMP mass, $\sigma_p$ is the only free
parameter and its limits are easily calculated for a given experiment.
These limits are routinely calculated by the dark matter experiments
and it has been found that the DAMA and CDMS results are incompatible
for most WIMP mass in the spin-independent case, assuming the standard
isothermal halo \cite{Akerib:2004fq} (an exception is the case of light
WIMPs of 6-9 GeV discussed in \cite{Gelmini:2004gm}).

\section{\label{sec:SDCS} Spin-Dependent Cross-Section}

For the remainder of the paper, we will focus on spin-dependent
interactions.  The generic form for the spin-dependent WIMP-nucleus
cross-section includes two couplings-- the WIMP-proton coupling $a_p$
and the WIMP-neutron coupling $a_n$,
\begin{eqnarray} \label{eqn:SDCS}
 \sigma_{SD}(q)
    &=& \frac{32 \mu^2 G_F^2}{2 J + 1}
        \left[a_p^2 S_{pp}(q) + a_p a_n S_{pn}(q) \right. \nonumber\\
    & & \qquad \qquad \qquad \left. + a_n^2 S_{nn}(q) \right] .
\end{eqnarray}
Here the nuclear structure functions $S_{pp}(q)$, $S_{nn}(q)$, and
$S_{pn}(q)$ are functions of the exchange momentum $q$ and are specific
to each nucleus.  The quantities $a_p$ and $a_n$ are actually the axial
four-fermion WIMP-nucleon couplings in units of $2\sqrt{2} G_F$
\cite{Gondolo:1996qw,Tovey:2000mm,darksusy}.

For a given experiment and WIMP mass, one can integrate
Eqn.~(\ref{eqn:dRdE}) over an energy bin $E_1$--$E_2$,  using the above
cross-section and factoring in efficiencies, quenching factors, etc.,
to obtain the expected number of recoil events $N_{rec}$ as a function
of  the parameters $a_p$ and $a_n$:
\begin{subequations} \label{eqn:conic}
\begin{equation} \label{eqn:conicrec}
 N_{rec}(a_p,a_n) = A_{rec} a_p^2 + B_{rec} a_p a_n + C_{rec} a_n^2 .
\end{equation}
$A_{rec}$, $B_{rec}$, and $C_{rec}$ are constants that can be
calculated from the integration (see Appendix). They differ between
experiments and they depend on the WIMP mass and WIMP velocity
distribution.  The expected value of the amplitude of the annual
modulation can be similarly calculated as Eqn.~(\ref{eqn:conicrec}):
\begin{equation} \label{eqn:conicmod}
 N_{ma}(a_p,a_n) = A_{ma} a_p^2 + B_{ma} a_p a_n + C_{ma} a_n^2
\end{equation}
\end{subequations}
where $N_{ma}$ is the modulation amplitude and $A_{ma}$, $B_{ma}$, and
$C_{ma}$ will again be calculable strictly from the experimental
parameters and the WIMP mass and velocity distribution.  A more
detailed discussion of calculating $N_{rec}$, $N_{ma}$, and the
associated constants is given in the Appendix.  Null search results
place an upper limit on $N_{rec}$ ($N_{rec} < N_{rec,max}$), which then
constrains the possible values for $a_p$ and $a_n$.  A positive
modulation signal, such as in DAMA ($N_{ma,min} < N_{ma} <
N_{ma,max}$), likewise constrains the possible values for $a_p$ and
$a_n$.

Previous authors, when searching for regions compatible with both the
DAMA signal and the null results of other experiments
\cite{Ullio:2000bv}, have studied the following two specific cases:
(i) $a_n = 0$ so that only protons contribute to WIMP interactions, and
(ii) $a_p = 0$ so that only neutrons contribute.  Here we examine the
more general case where both $a_p$ and $a_n$ may be non-zero (some of
this parameter space has also been examined by Giuliani
\cite{Giuliani:2004uk}).  To do this, we examine the form of the limits
on $a_p$ and $a_n$ implied by Eqn.~(\ref{eqn:conic}).

Eqn.~(\ref{eqn:conic}) describes a conic section in the $a_p$-$a_n$
plane. This conic section can be an ellipse, a hyperbola, or a set of
two straight lines (it cannot be a parabola because linear terms in
$a_p$ or $a_n$ are absent). Which is the case depends on the values of
the coefficients $A$, $B$, and $C$. For the benefit of the reader, we
will briefly review the relevant shapes here (our conics are always
centered at the origin, so we only describe such cases below).  

An ellipse in the $x$-$y$ plane can be written as:
\begin{equation} \label{eqn:ellipse}
 \frac{x^2}{a^2} + \frac{y^2}{b^2} = 1
\end{equation}
This ellipse is symmetric about the $x$ and $y$ axes.  Adding a
cross-term (i.e. $x y$) in Eqn.~(\ref{eqn:ellipse}) (like the $B$ term
in Eqn.~(\ref{eqn:conic})) essentially rotates the ellipse- the major
and minor axes do not lie on the $x$ and $y$ axes, but along some new
rotated axes.

Another conic section is the hyperbola:
\begin{equation} \label{eqn:hyperbola}
 \frac{x^2}{a^2} - \frac{y^2}{b^2} = 1
\end{equation}
which, in this form, gives two branches of a hyperbola open along the
$+x$ and $-x$ directions.  Rotation of this form also leads to a
cross-term.

Two parallel lines are also a conic (technically, a ``degenerate''
conic), that can be written in the form, e.g.:
\begin{equation} \label{eqn:parallel}
 x^2 = a^2
\end{equation}
which describes two lines parallel to the $y$-axis, located at
$x = +a$ and $x = -a$.  Other parallel lines can be rotated into
this form as well.

Note that, while other conics exist besides the three described, these
are the only forms possible for Eqn.~(\ref{eqn:conic}) (which is not
the most general second degree polynomial).


\begin{table*}
 \begin{tabular}{||l|l|l|l|l|c||}
 \hline\hline
 Experiment & Exposure [kg-day] & Threshold [keV] & Efficiency [\%]
    & Constraint (for stated recoil energies) & Ref. \\
 \hline
     CDMS I
   & Si: 6.58
   & 5
   & \multirow{3}{*}[-1.0ex]{\begin{minipage}[c]{75pt} \begin{flushleft}
                    $E<10$ keV: 7.6 \\ $E<20$ keV: 22.8 \\ $E>20$ keV: 38
                    \end{flushleft} \end{minipage}}
   & 5--55 keV: $<$2.3 events ($\dagger$)
   & \protect\cite{Akerib:2003px} \\
 \cline{1-3} \cline{5-6}
     CDMS II
   & Ge: 52.6
   & 10
   &
   & 10--100 keV: $<$2.3 events ($\dagger$)
   & \protect\cite{Akerib:2004fq} \\   
 \cline{1-3} \cline{5-6}
     \begin{minipage}{50pt} \begin{flushleft}
     CDMS II\\(projection)
     \end{flushleft} \end{minipage}
   & \begin{minipage}{50pt} \begin{flushleft}
     Si: 300\\Ge: 1200
     \end{flushleft} \end{minipage}
   & 5
   &
   & 5--100 keV: $<$2.3 events ($\dagger$)
   & \protect\cite{cdms:projection} \\
 \hline
     EDELWEISS
   & Ge: 8.2
   & 20
   & 100
   & 20--100 keV: $<$2.3 events  ($\dagger$)
   & \protect\cite{Sanglard:2003ht} \\
 \hline
     CRESST I
   & Al$_2$O$_3$: 1.51
   & 0.6
   & 100
   & ($\ddag$)
   & \protect\cite{Angloher:in} \\
 \hline
     CRESST II
   & CaWO$_4$: 10.448
   & 10
   & 100
   & \begin{minipage}{180pt} \begin{flushleft}
     Ca+O, 15--40 keV: $<$6 events  \\ W, 12--40 keV: $<$2.3 events ($\dagger$)
     \end{flushleft} \end{minipage}
   & \protect\cite{Stodolsky:2004dsu} \\
 \hline
     DAMA/Xe-2
   & Xe: 1763.2
   & 30 ($\circ$)
   & 100
   & ($\ddag$)
   & \protect\cite{Bernabei:ad} \\
 \hline
     DAMA/NaI
   & NaI: 107731
   & \begin{minipage}{50pt} \begin{flushleft}
     I: 22 ($\diamond$) \\ Na: 6.7 ($\diamond$)
     \end{flushleft} \end{minipage}
   & 100
   & 2--6 keVee: 0.0200$\pm$0.0032/kg-day-keVee ($\bullet$)
   & \protect\cite{Bernabei:2003za} \\
 \hline
     Elegant V
   & NaI: 111854
   & \begin{minipage}{50pt} \begin{flushleft}
     I: 44 ($\diamond$) \\ Na: 13.4 ($\diamond$)
     \end{flushleft} \end{minipage}
   & 100
   & 4--5 keVee: 0.009$\pm$0.019/kg-day-keVee ($\bullet$)
   & \protect\cite{Yoshida:2000vk} \\
 \hline\hline
 \end{tabular} 
 \caption{
   Experimental constraints used in this study.
   Notes to the table:
    ($\dagger$) upper limit assuming no detected event; 
    ($\ddag$) limits generated based upon the binned data available
      in the corresponding papers (does not include any background
      subtraction and may be conservative),;
    ($\circ$) from an electron equivalent threshold of 13~keVee,
      using the quenching factor $Q=E_{\rm ee}/E$ equal to 0.44 for Xe
      \protect\cite{Bernabei:ad};
    ($\diamond$) from an electron equivalent thresholds of 2~keVee
      (DAMA/NaI) and 4~keVee (Elegant V),
      using $Q$ equal to 0.09 for I and  0.3 for Na
      \protect\cite{Bernabei:2003za};
    ($\bullet$) amplitude of annual modulation.
 }
 \vspace{-5pt}
 \label{tab:ExpParam}
\end{table*}

Eqn.~(\ref{eqn:conic}) can, under suitable rotation in the $a_p$-$a_n$
plane, be expressed in one of the forms in Eqns.~(\ref{eqn:ellipse}),
(\ref{eqn:hyperbola}), or (\ref{eqn:parallel}).  The shape of our conic
can be determined by the value of the discriminant 
\begin{equation} \label{eqn:conicGamma}
 \gamma \equiv B^2 - 4 A C,
\end{equation}
as follows: $\gamma < 0$ corresponds to an ellipse, $\gamma > 0$ to a
hyperbola, and $\gamma = 0$ to two parallel lines.  Due to our
conics always being centered at the origin, there is a symmetry under
$(a_p,a_n) \to (-a_p,-a_n)$.

Null result experiments provide an upper bound $N_{max}$ on the number
of recoils or modulation amplitude.  This upper bound restricts the
allowed $a_p$-$a_n$ parameter space to the region inside the ellipse
defined by $N_{max}$ (for the case in which Eqn.~(\ref{eqn:conic})
describes an ellipse), or to the region containing the origin between
the two branches of the hyberbola or the two parallel lines defined by
$N_{max}$ (for the case in which Eqn.~(\ref{eqn:conic}) describes a
hyperbola or two straight lines).  Couplings outside these regions
would lead to more recoils than observed; couplings within these
regions yield less recoils than $N_{max}$.  

Experiments which obtain positive signals (such as claimed by DAMA)
restrict the $a_p$-$a_n$ parameter space to a band in the shape of the
appropriate conic.  The width of the band is defined by the uncertainty
in the observed signal (the uncertainty and hence the width of the band
can be given to various degrees of accuracy, e.g.\ 1 or 2$\sigma$).
For example, if the conic is a circle, the allowed region would be an
annulus; regions ``outside'' this annulus would give either too small
or too large a signal.  

The relative magnitudes of $A$, $B$, and $C$ in Eqn.~(\ref{eqn:conic})
are primarily dependent upon the structure functions of
Eqn.~(\ref{eqn:SDCS}).  At zero recoil energy, the structure functions
in Eqn.~(\ref{eqn:conic}) for a single nuclear element form a complete
square, of the form $N\propto (\alpha a_p + \beta a_n)^2$ (see,
e.g., Eqn.~(13) of Ref.~\cite{Engel:wq}). This is a rotation of
Eqn.~(\ref{eqn:parallel}). Hence for $q = 0$, we have $\gamma = 0$ and
the appropriate conics are two parallel lines.
However, at nonzero momentum transfer, interference and spin effects
destroy the symmetry so that different conics may result.  In addition,
detectors with multiple elements, such as the NaI of DAMA, have
structure functions from all of these elements contributing to the
coefficients in Eqn.~(\ref{eqn:conic}); the resulting conic may  be
very different to that obtained from the individual elements (see e.g.\ 
Figure \ref{fig:cdmsEllipses}).  For the experiments studied in this
paper, we find that the relevant conics are either two parallel lines
or ellipses.  Some of the ellipses are so elongated as to appear as two
parallel lines in our plots.

We will examine the allowed shapes in the $a_p$-$a_n$ plane for current
experiments and then determine where they overlap. In this way we will
find those regions in parameter space that are consistent with all the
current experimental results.


\begin{figure*}[t]
 \includegraphics[width=1.0\textwidth]{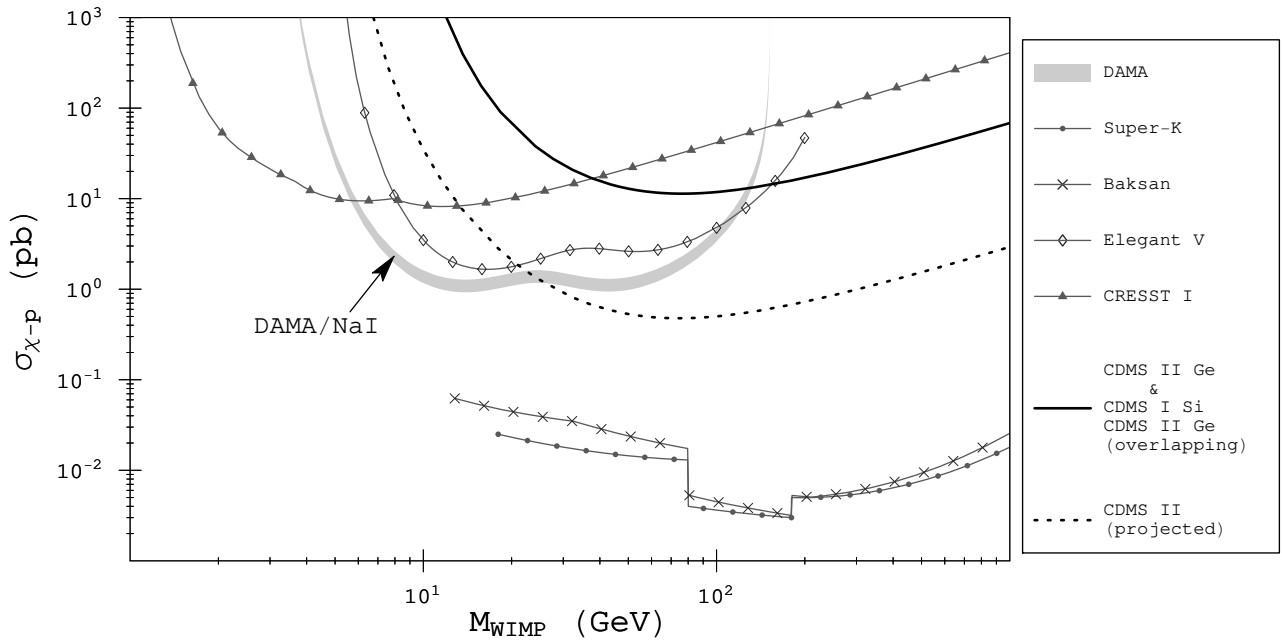}
 \caption[Neutralino-proton cross-section limits]{
   WIMP-proton cross-section limits for the case $a_n = 0$.
   The CDMS Si+Ge limit overlaps that of Ge only; the Si
   is relatively insensitive in this case.
   Super-K only analyzed their data for WIMP masses above 18 GeV;
   their limit is taken from Figure 14 of \cite{Desai:2004pq}.
   Baksan analyzed fluxes down to 13 GeV.
   Super-K and Baksan rule out the DAMA results over their analysis
   ranges and CRESST I limits DAMA at low masses, but WIMPs between
   6-13 GeV are consistent with all results for this case.
   }
 \label{fig:protonCS}
\end{figure*}


\begin{figure*}[t]
 \includegraphics[width=1.0\textwidth]{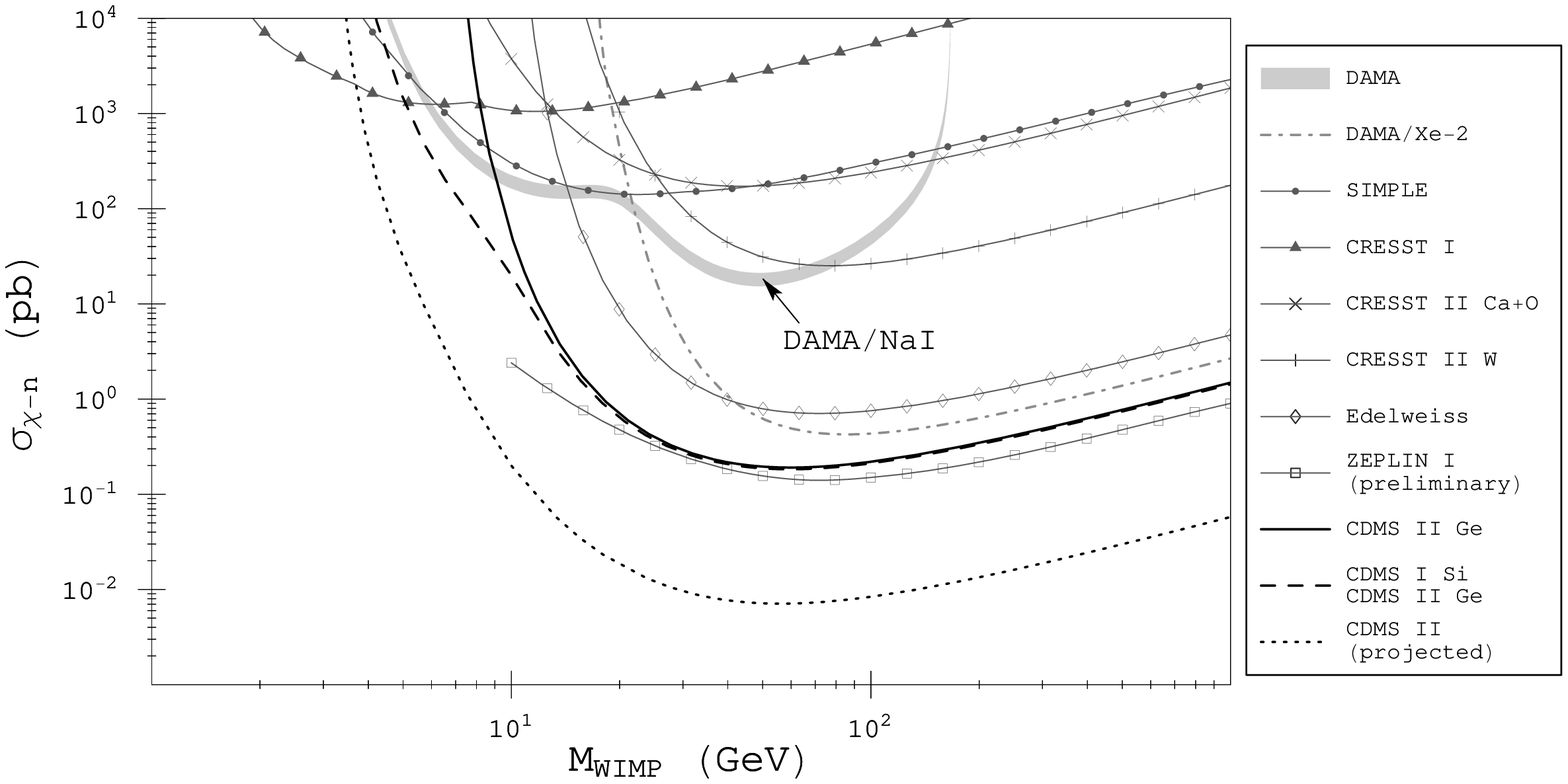}
 \caption[Neutralino-neutron cross-section limits]{
   WIMP-neutron cross-section limits for the case $a_p = 0$.
   SIMPLE limits are taken from Figure 1 of \cite{Giuliani:2004uk}.
   The addition of the Si data clearly benefits CDMS at lower masses,
   ruling out DAMA for this simple case.
   }
 \label{fig:neutronCS}
\end{figure*}

\section{\label{sec:Experiments} Experiments}

The constraints from different experiments on the $a_p$-$a_n$ phase
space are highly dependent on the choice of detector materials.  Odd Z
elements, such as Sodium, Iodine, and Hydrogen, are more sensitive to
$a_p$ and thus generate narrow ellipses with the semi-minor axis close
to the $a_p$ axis (interference effects can cause the ellipse to be
rotated slightly).  Conversely, elements with an odd number of
neutrons, such as Xenon-129, Silicon-29, and Germanium-73, are more
sensitive to $a_n$ and their associated ellipses have semi-minor axes
near the $a_n$ axis.  Spinless elements such as most isotopes of
Silicon and Germanium (used by many experiments looking for
spin-independent couplings), provide essentially no constraint for
either $a_p$ or $a_n$.

Table \ref{tab:ExpParam} (extended from \cite{Gelmini:2004gm}) displays
the various experimental constraints used to generate our results.
In Figures \ref{fig:protonCS} \& \ref{fig:neutronCS}, we show the
limits from various experiments on the WIMP-proton and WIMP-neutron
allowed couplings for the cases $a_n = 0$ and $a_p = 0$, respectively.
Due to the novel detection technique used by SIMPLE (and the
additional complexities involved) \cite{Collar:2000fi}, we did not
re-analyze their data in our study, but include in Figure
\ref{fig:neutronCS} the WIMP-neutron cross-section presented in
Figure 1 of \cite{Giuliani:2004uk}.  The full ZEPLIN I results were not
available during the writing of this paper so were not included in our
analysis; however, a preliminary limit presented by the ZEPLIN group is
also included in Figure \ref{fig:neutronCS} \cite{idm2004:zeplin}.

Several experiments providing recent results, including Elegant V
(NaI), CRESST I (Al), SIMPLE (F), and DAMA/NaI, used odd Z nuclei and
were thus more sensitive to the WIMP-proton coupling $a_p$. Other
experiments, including DAMA/Xe-2 (Xe-129), CRESST II (Ca-43, W-183,
and O-17), Edelweiss (Ge-73), ZEPLIN I (Xe-129 and Xe-131), and CDMS
(Si-29 and Ge-73) used neutron odd nuclei and were more sensitive to
the WIMP-neutron coupling $a_n$.  We note, however, that each of these
experiments has some sensitivity to both $a_p$ and $a_n$, as the minor
axes of their ellipses are \textit{not exactly aligned} with either
$a_p$ or $a_n$.  Hence, surprisingly, CRESST and SIMPLE obtain very
good results on WIMP-neutron coupling due to their low thresholds.


The only experiment to claim a result, DAMA/NaI, uses odd Z NaI
detectors and a large exposure to search for an annual modulation.
They observe an annual modulation amplitude of $0.0200 \pm 0.0032$
events/kg/day/keVee over electron-equivalent recoil energies of 2-6
keV.  Such a modulation has been shown to be generally inconsistent with
other experiments for spin-independent interactions, but we shall see
there are choices of parameters still allowed for the spin-dependent
case.  The DAMA limits we place in $a_p$-$a_n$ space are those that
reproduce the above modulation amplitude, using the structure functions
given by Ressell and Dean \cite{Ressell:1997kx}.  The WIMP cross
sections on protons and neutrons implied by the DAMA result are plotted
in Figures  \ref{fig:protonCS} \& \ref{fig:neutronCS} respectively.
Kinematics alone can restrict the allowed masses of this modulation:
small masses do not generate enough recoils above threshhold.  One
can see this in the figures as DAMA begins to lose sensitivity below
about 4 GeV, requiring large cross-sections to account for the observed
signal.  This feature is common to all dark matter experiments and
lower threshold energies are required to detect the smaller recoil
energies of low masses.  This lower limit on sensitivity is also
dependent on the parameters of the velocity distribution and may
change significantly in the presence of streams \cite{Gelmini:2004gm}.
Larger masses are also ruled out as the phase of the modulation
would be reversed.  For the 2-6 keV data plotted, the limit is roughly
180 GeV; however, this limit is dependent upon the range of recoil
energies.  Freese and Lewis have examined this phenomenon and
determined that masses above 103 GeV would reverse the modulation at
2 keV for DAMA \cite{Lewis:2003bv}.  In addition to the 2-6 keV data,
DAMA released modulation amplitudes for slightly different recoil bins,
namely for recoil energies 2-4 keV ($0.0233 \pm 0.0047$ /kg/day/keVee)
and 2-5 keV ($0.0210 \pm 0.0038$ /kg/day/keVee).  At the masses we are
considering ($< 100$ GeV), the results of this paper are the same for
any of the three DAMA binnings.


All other experiments have null results which place bounds on the
couplings.  Elegant V used NaI detectors to search for an annual
modulation and found no counts above the statistically limiting
background \cite{Yoshida:2000vk}.  While providing one of the best
exclusion limits for a proton-only coupling at low WIMP masses, the
sensitivity of this experiment is not enough to examine the DAMA
observed region (see Figure \ref{fig:protonCS}).  CRESST I
\cite{Angloher:in} and SIMPLE \cite{Collar:2000fi} provide similar
limits on WIMP-proton coupling.  As mentioned above, due to their very
low threshholds, CRESST I and SIMPLE also provide what have been the
best neutron-only limits at low WIMP masses.  DAMA/Xe-2, a neutron-odd
experiment bounds the WIMP-neutron cross-section above 40 GeV
\cite{Bernabei:ad}.  Our examination of Edelweiss (similar to CDMS,
described below), shows that they produce a similar limit to DAMA/Xe-2
\cite{Sanglard:2003ht} (see Figure \ref{fig:neutronCS}).


Super-Kamiokande, on the other hand, is an indirect detection
experiment, searching for high-energy neutrinos produced by the
annihilation of WIMPs in the Sun's core after being gravitationally
captured.  Since the Sun is predominantly composed of hydrogen, the
capture rate (and thus neutrino flux) is particularly sensitive to the
WIMP-proton coupling $a_p$.  The Super-K detector searched
for this additional neutrino flux by observing upward moving muons
(induced by muon neutrinos) \cite{Desai:2004pq}.  The lack of an
additional muon signal leads to the most significant $a_p$ bounds
above the analysis threshold WIMP mass of 18 GeV; below this mass, a
significant portion of the muons would stop in the detector, rather
than pass through, making the data more difficult to analyze.
Discontinuities of the Super-K published limits occur at 80 GeV (W
mass) and 174 GeV (top mass) due to different possible annihilation
products.  The Super-K WIMP-proton cross-section limits are shown in
Figure \ref{fig:protonCS}; these limits are taken directly from
Ref.~\cite{Desai:2004pq}, derived therein using the model-independent
technique of Kamionkowski \textit{et al.}~\cite{Kamionkowski:1994dp}.
We note that the bounds from Super-K rely on three assumptions: First,
we assume that the WIMPs have achieved equilibrium in the Sun, so that
the capture rate is equal to the annihilation rate; this assumption is
almost certainly correct, particularly at small WIMP masses.  Second,
we assume that either WIMPs are equal to their antipartners so that
they can annihilate among themselves or, if WIMPs are not the same as
the antiWIMPs, that there is no WIMP/antiWIMP asymmetry.  Third, the
WIMPs do not decay to light fermions only (in which case fewer
observable neutrinos would be produced).  The second and third ways to
evade the SuperK bounds were pointed out by Kurylov and Kamionkowski
\cite{Kurylov:2003ra}.  Here we assume that all three assumptions are
correct; then Super-K rules out the proton only coupling for
WIMP masses above 18 GeV.

Baksan, another indirect detection experiment, provides the tightest
bounds on proton only coupling between 13 and 18 GeV.  Although Baksan
provides weaker WIMP-induced muon flux limits than Super-K above 18
GeV, the Baksan experiment has analyzed these fluxes down to a lower
WIMP mass of 13 GeV \cite{bib:baksan}.  These flux limits can be used
to determine WIMP-proton cross-section limits in the same manner as
Super-K; such an analysis is not available from the Baksan
collaboration at this time.  However, the Baksan cross-section limit
can be determined by rescaling the Super-K limit by the relative flux
limits of these two experiments (since the expected flux is
proportional to this cross-section).  Below 18 GeV, where there is no
Super-K limit to rescale, we extrapolate the Baksan cross-section limit
using their given flux limit and the WIMP mass dependence of the
expected flux, given by Eqn.~(9.55) of Ref.~\cite{Jungman:1995df}.  We
find that Baksan rules out proton only coupling down to WIMP masses of
13 GeV.


CDMS II, in the Soudan Underground Laboratory, uses Silicon and
Germanium detectors with strong discrimination to provide the
strongest spin-independent cross-section limits to date
\cite{Akerib:2004fq}.  Analysis of CDMS results in the spin-dependent
sector have consistently been ignored, as the small natural abundance
of spin odd isotopes in Silicon (4.68\% Si-29) and Germanium (7.73\%
Ge-73) has given the impression that CDMS is not sensitive enough for
any significant spin-dependent results.  Our analysis, however, shows
that the extremely clean, background-free data is sufficient to
overcome the smaller spin sensitive mass, with the odd neutron Ge-73
(Silicon data has not yet been published) providing significant limits
to $a_n$.  The limits we obtain are due to the fact that no events
were observed in 52.6 days of livetime for four 0.250 kg Germanium
detectors over 10 keV to 100 keV recoil energy, using an efficiency of
0.228 for 10-20 keV and 0.38 for 20-100 keV; an efficiency of 0.076
for 5-10 keV is used to determine limits at lower thresholds.  Since
Silicon is lighter than Germanium, it will provide better limits at
small WIMP masses.  As the Silicon data has not yet been published for
the first run of CDMS II, we augment the CDMS II Ge results with the
most recent CDMS I Si data \cite{Akerib:2003px}.  The last CDMS I run
used the same detector tower as now used in CDMS II, containing two
0.100 kg Silicon detectors as well as the four Germanium ones.
However, one of these Si detectors was contaminated and not used in the
data analysis.  No Si events were found in the range 5-55 keV after
65.8 days of livetime, with the same efficiencies as above.  While
several Si events were observed above 55 keV, these are consistent with
a neutron background.  Even if these events were WIMP scatters, they are
not consistent with a light mass.  Since the Si data is being used to
extend the limits at low masses (Ge data dominates at higher masses),
we feel justified in ignoring the high energy events.  CDMS hopes to
install a total of five towers and achieve a total exposure during the
CDMS II run of 1200 kg days for Ge and 300 kg days for Si; we will use
these numbers to project future limits, assuming a 5 keV threshold and
null results \cite{cdms:projection}.  We have used the structure
functions given by Ressell \textit{et al.}~\cite{Ressell:1993qm} for
Si-29 and Dimitrov \textit{et al.}~\cite{Dimitrov:1994gc} for Ge-73; a
comparison of the Ge-73 functions from Ressell shows the results are
relatively insensitive to the models used to derive the form factors.


\begin{figure*}
 \includegraphics[width=0.575\textwidth,trim=0 0 0 7]{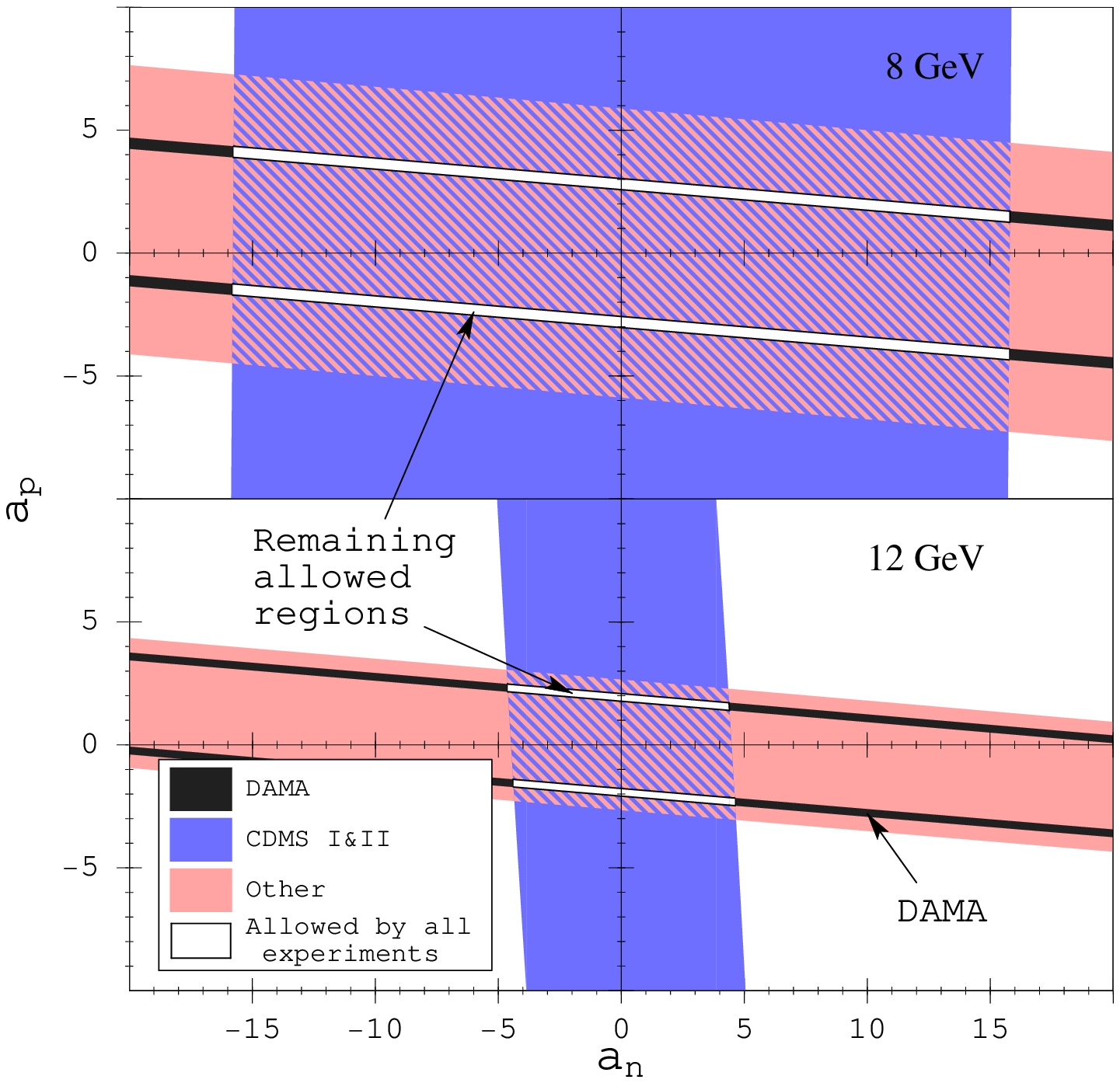}
 \hfill
 \includegraphics[width=0.40\textwidth,trim=0 7 0 17]{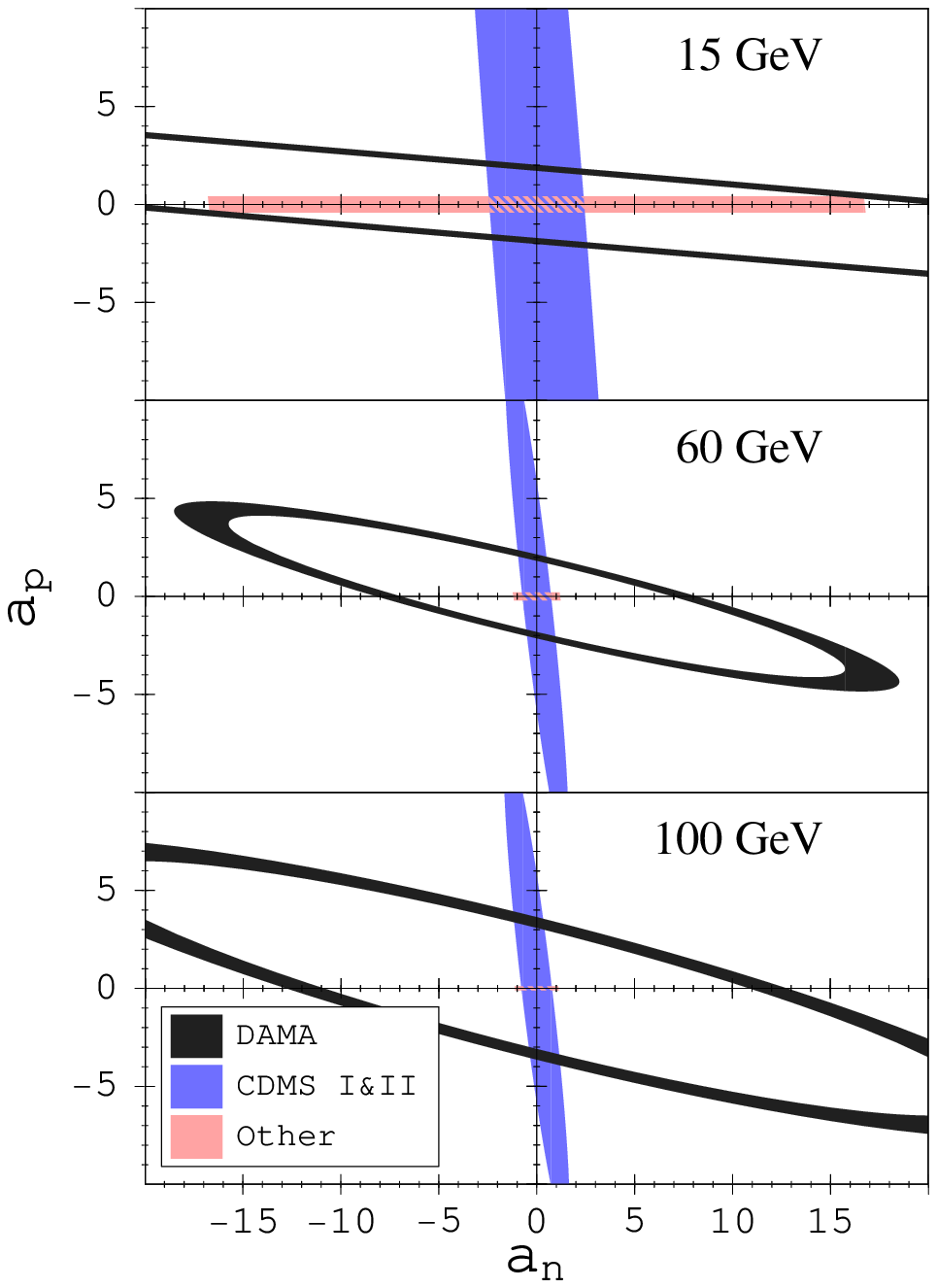}
 \caption[SD allowed couplings]{
   Spin-dependent allowed couplings at several different WIMP
   masses (in different panels).  The legends in the figure indicate
   regions allowed by different experiments. The thin black horizontal
   bands (ellipses at higher masses) show the region allowed by DAMA
   (note that the black bands are indicated as white for regions
   allowed by all experiments); the allowed couplings for DAMA are
   those that fall \textit{on} the bands, not between them. The dark
   (blue) vertical regions correspond to couplings allowed by the CDMS
   I Si and CDMS II Ge data sets.  The remaining (pink) horizontal
   region, labelled as ``other'' in the legend, represents couplings
   allowed by the combined null results of Super-K, Baksan, Elegant V,
   CRESST, DAMA/Xe-2, and Edelweiss (couplings outside of this region
   are ruled out by at least one of these experiments).  The region
   allowed by \textit{all} experiments is indicated by white bands.
   Only inside these white bands is the DAMA signal consistent with the
   null results from other experiments.  For masses above 13 GeV there
   is no consistent region (no white band).
   }
 \label{fig:couplingArray}
\end{figure*}

\section{\label{sec:Results} Results}

We first present results for the simple cases where the WIMP couples to
only the proton or neutron.  From a combination of constraints in
Figure \ref{fig:neutronCS}, one can see that DAMA results are not
compatible with limits from other experiments for the case of the WIMP
coupling with neutrons only.  However, one can see in
Figure \ref{fig:protonCS} that DAMA still survives for the case of
proton only coupling for masses less than 18 GeV.

Figure \ref{fig:protonCS} displays the current limits for the
WIMP-proton cross-section assuming the WIMP couples only to the proton
($a_n = 0$).  CDMS does not currently constrain this coupling.
The Super-K and Baksan results, however, strongly rule out the
WIMP-proton coupling that would be required to explain DAMA for masses
above 13 GeV.  Elegant V still allows for the DAMA observed modulation
at these lower masses and CRESST I provides only a lower mass limit of
6 GeV.

Figure \ref{fig:neutronCS} displays the WIMP-neutron cross-section in
the alternative case that the WIMP couples only to the neutron $a_p =
0$.  DAMA/Xe-2 and Edelweiss rule out this case for masses above 20 GeV.
The surprising result is that CDMS II significantly improves upon this
limit, ruling out masses above 10 GeV.  Furthermore, if one includes
the CDMS I Si data, CDMS rules out all of the DAMA allowed region for
this case.  Recent ZEPLIN results are extremely powerful as well.

While the above two cases demonstrate the different sensitivities
of the experiments, they are not necessarily representative of the
more general parameter space, where \textit{both} $a_p$ and $a_n$ may
be non-zero.  Figure \ref{fig:couplingArray} shows the $a_p$-$a_n$
regions allowed by DAMA, CDMS, and a combination of the remaining
experiments at several different masses for this more general case.

The DAMA observed modulation generates an elliptical ring, flattened
in roughly the direction of the $a_p$-axis (due to the odd Z NaI).
[At low masses, the portion of the elliptical ring that is plotted looks
like two parallel bands.]  The ellipse is slightly rotated from the
$a_p$-$a_n$ axes; this rotation increases at larger masses as observed
recoils tend to come from the Iodine atoms rather than the lighter
Sodium (which has different structure functions).  The DAMA allowed
couplings are those that fall \textit{on} the elliptical bands, not
between them; couplings ``inside'' the inner edge of the elliptical
ring (between the two bands of the ellipses) are too small to generate
the observed modulation.

Other experiments, which have null results, place bounds on the
$a_p$-$a_n$ parameter space, as shown in Figure \ref{fig:couplingArray}.
The light grey (pink) horizontal regions, labelled as ``other'' in the
legend of the figure, illustrate a combination of limits from Super-K,
Baksan, Elegant V, CRESST, Edelweiss, and DAMA/Xe-2 (we remind
the reader that DAMA/Xe-2 is entirely different from the DAMA/NaI
annual modulation results). Only couplings within these regions
satisfy the null result constraints of all these experiments.
Super-K and Baksan results depend upon WIMPs interacting with
hydrogen, and thus inherently limit the proton coupling $a_p$ only.
The Super-K and Baksan allowed regions are bands along the
$a_n$-axis (corresponding to the region between two parallel lines).
The DAMA/Xe-2 and Edelweiss allowed ellipses, whose minor axes are
along the $a_n$ direction, are far broader in the $a_p$ direction.
The intersection of these regions (horizontal Super-K \& Baksan
bands with vertical DAMA/Xe-2 \& Edelweiss ellipses) forms a
roughly rectangular region above 13 GeV, clearly visible in the 15 GeV
panel and getting smaller at higher masses.  Any allowed model must
lie within this ``rectangle.''  Below 18 GeV, Super-K and DAMA/Xe-2 no
longer provide the best limits (or any limit, in Super-K's case);
below 13 GeV, Baksan and Edelweiss also lose their sensitivity.
For these lower masses, Elegant V and CRESST I (included in the region
labelled ``other'') provide the best limits; however, they do not
significantly constrain the DAMA space between 5 and 13 GeV.

The recent CDMS results are also shown in Figure
\ref{fig:couplingArray}.  CDMS's null result, based upon odd neutron
Si and Ge, generates an allowed region elongated along
the $a_p$-axis. In Figure \ref{fig:couplingArray}, the region allowed
by the CDMS data is dark grey (blue) and is labelled ``CDMS I\&II.''
The addition of CDMS I Si data increases sensitivity at low masses,
due both to the lighter mass and lower threshhold than the CDMS II Ge
data.  As the sensitivity changes from predominantly Ge at high masses
to Si at low masses, the shape of the allowed region rotates slightly
(as Si and Ge have different structure functions).

In sum, we have applied null results from all other experiments to see
whether or not DAMA annual modulation results can be compatible.  The
only remaining regions are shown in white bands in Figure
\ref{fig:couplingArray}.  Only WIMP masses in the range 5-13 GeV
survive as spin-dependent candidates matching all the data.
In the next section (Summary and Discussion) we discuss the viable
ranges of couplings in the $a_p$-$a_n$ plane.

\begin{figure}
 \includegraphics[width=1\columnwidth]{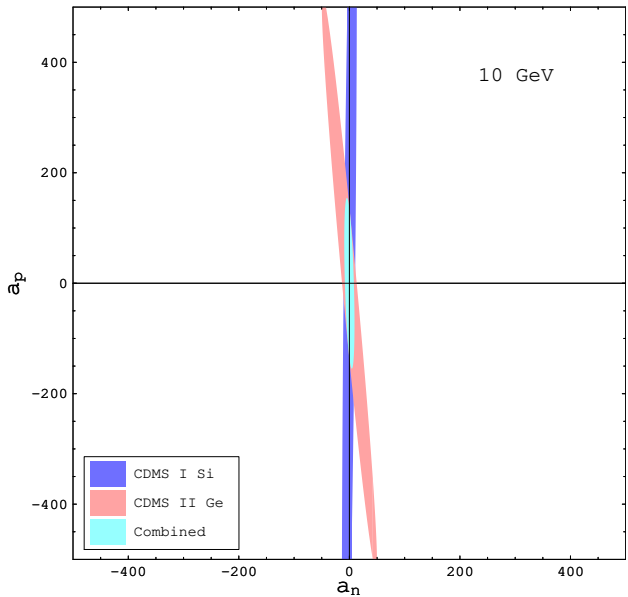}
 \caption[CDMS Si and Ge Limits at 10 GeV]{
   Spin-dependent couplings allowed by CDMS at a WIMP mass of 10 GeV.
   The separate limits from Si and Ge each produce extremely long
   ellipses; the combined limit is significantly smaller.
   }
 \label{fig:cdmsEllipses}
\end{figure}

Note that, when one allows both $a_p$ and $a_n$ to be nonzero, the
bounds of Figures \ref{fig:protonCS} \& \ref{fig:neutronCS} on
proton-only and neutron-only couplings can be considerably weakened.
We illustrate this with an example in Figure \ref{fig:cdmsEllipses},
which shows the limits in the $a_p$-$a_n$ plane separately for the
CDMS I Si and CDMS II Ge at a WIMP mass of 10 GeV.  While Ge limits
$a_p$ to below 160 for $a_n = 0$, even a small non-zero value of $a_n$
allows for $a_p$ values several times larger (e.g.\  we can have
$a_p = 330$ at $a_n = -20$).  The CDMS II Ge limit in Figure
\ref{fig:protonCS} for the WIMP-proton coupling applies only to the
case of $a_n = 0$, but the limit in the general case is more than an
order of magnitude weaker.


Referring again to Figure \ref{fig:cdmsEllipses}, the Si data forms
a similar elliptical limit as the Ge, but slightly rotated.  Even the
slight rotation is significant enough that the two narrow ellipses
rarely overlap- the combined Si and Ge limit is significantly smaller
than that of either of the separate limits.  Detectors employing
multiple elements can break the (near) degeneracy typical of individual
elements.

\section{\label{sec:Discussion} Summary and Discussion}


\begin{figure}
 \includegraphics[width=1\columnwidth]{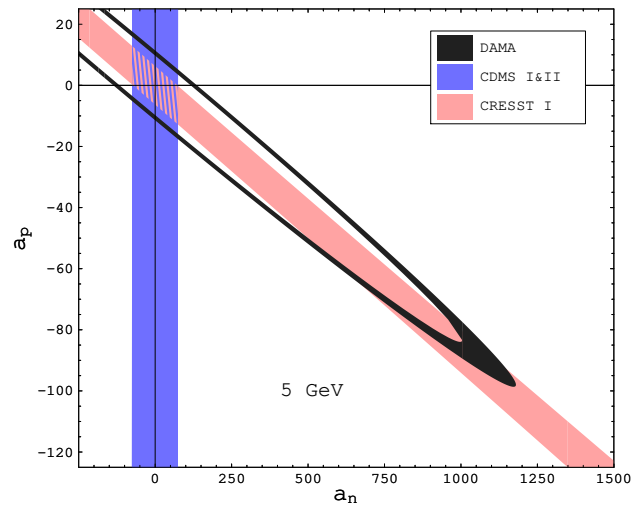}
 \caption[SD Allowed Couplings at 5 GeV]{
   Spin-dependent allowed couplings at 5 GeV for DAMA (black elliptical
   band), CDMS (vertical region), and CRESST I (angled region).
   The CDMS and CRESST I null results only allow for couplings within
   the intersection of their corresponding allowed regions (hatched
   region);
   this intersection does not include any couplings that can
   reproduce the DAMA observed modulation.
   Below 5 GeV, the CDMS/CRESST I constraint grows rapidly stronger
   relative to the DAMA required couplings.
   Note the axes are not at the same scale.
   }
 \label{fig:lowMass}
\end{figure}

Figure \ref{fig:couplingArray} illustrates our main results.  When we
allow both $a_p$ and $a_n$ to be nonzero,  there is no region of
parameter space with WIMP masses above 13 GeV which is compatible with
both the positive results of DAMA and the null results of Super-K,
DAMA/Xe-2, Edelweiss, and CDMS.
Below 13 GeV, there are line segments in the $a_p$-$a_n$ plane
which are still compatible with the DAMA data and all null results
from other experiments; CDMS II is the most constraining.
At 12 GeV, the compatible line segments are $a_p = -0.084 a_n + 1.94
\, (\pm 0.16)$ with $-4.62 \leq a_n \leq 4.39$; at 8 GeV they are
$a_p = -0.084 a_n + 2.81 \, (\pm 0.23)$ over $-15.8 \leq a_n
\leq 15.8$ (these describe only one of the line segments, the other
is found by taking $(a_n,a_p) \to (-a_n,-a_p)$).  This includes cases
for which the proton and neutron couplings are comparable
(e.g.\ $a_p = a_n = 1.8$ at 12 GeV and $a_p = a_n = 2.6$ at 8 GeV),
in which case the WIMP-proton and WIMP-neutron cross-sections
are of the same order (equal for the given examples).
We also note that only models with $|a_p| > 1$ are consistent at
\textit{any} mass.  Below about 5 GeV, CRESST I becomes significant
enough to exclude (in combination with CDMS) all of DAMA's parameter
space (Figure \ref{fig:lowMass}).

We note that the small remaining regime of spin-dependent parameter
space could be confirmed or ruled out by more complete analysis of
Super-K or other solar neutrino (indirect detection) data.  If Super-K,
Baksan, and ANTARES (or in the future IceCube and ANTARES) were able to
push down their analyses below 13 GeV WIMP masses, they would be able to
either find the WIMP annihilation from the Sun compatible with DAMA, or
rule out DAMA entirely as due to spin-dependent interactions of a WIMP
that is its own anti-WIMP.  If the WIMP and anti-WIMP abundances in the
Sun would differ substantially, there would be no annihilation signal
from the Sun even if WIMPs may be abundant enough to give rise to a
signal in DAMA.  Previously Ullio \textit{et al.}~\cite{Ullio:2000bv}
attempted to push the Super-K results down to lower WIMP masses.
However, we feel that their analysis may have been too constraining due
to the fact that the energy threshold for the muons in the experiments
was not taken into account (it was set to zero).  We feel that this
analysis must be left to the experimentalists.  We encourage the
experimentalists to study the bounds at lower WIMP masses that would
arise when one extends the analysis to more difficult cases: muons that
stop in the detector, muons that start in the detector, and contained
events (muons that start and stop in the detector).  For the contained
events one should consider both muon and electron neutrinos. If such an
analysis were possible, then one could confirm or refute the remaining
parameter region for spin-dependent interactions in DAMA.

In the future, as other experiments become more sensitive, they will
of course further restrict the parameter space as well.  New CRESST I
or SIMPLE data should also be able to push down the WIMP mass
compatible with DAMA.  Edelweiss, CRESST II and ZEPLIN
\cite{Luscher:2001yn}, whose role is similar to CDMS in this regard
(their ellipses align with the $a_p$ axis), will also become important.

Accelerator bounds on spin-dependent interactions are far weaker than
the limits we have discussed from dark matter experiments.
At the high energies in accelerators, the WIMPs couple directly to the
quarks and can no longer be treated as effectively coupling to the
nucleons as a whole.  Such scattering may occur through exchange of
several different particles, leading to interference effects that
further reduce the cross-section.  Likely signatures of such scatters
are also either unobservable or dwarfed by backgrounds.  Looking at the
case of a simple $Z$ exchange, the lowest order case of $q \bar{q}
\rightarrow Z + X$ with $Z$ producing a WIMP pair $Z \rightarrow \chi
\chi$ would not present an observable signal as the WIMP itself, being
electrically neutral and weakly interacting, is not directly detectable.
The signal is potentially observable if a gluon is also emitted, 
$q \bar{q} \rightarrow Z + g + X$, but other Standard Model processes
produce the same signatures at rates much larger than those expected
for $a_p$ and $a_n$ of order one.  Even assuming no interference
effects, the experimental limits from this and similar processes give
$a_p \lae O(10^2)$, much weaker than the limits of $O(1)$ from the dark
matter experiments.  As a final remark, we note that the the Minimal
Supersymmetric extension of the Standard Model does not provide a dark
matter candidate with the characteristics pointed to by our study.


\begin{acknowledgments}
  C. Savage thanks G. Kane and R. Schnee for useful discussions.
  We thank M. Kamionkowski for kind comments on the first version of
  this paper.
  We acknowledge the support of the DOE and the Michigan Center for
  Theoretical Physics via the University of Michigan.
  We thank the MCTP for hosting the Dark Side of the Universe Workshop,
  during which most of this work was performed.
\end{acknowledgments}


\appendix*

\section{\label{sec:DetRec} Detector Recoils}

To determine the number of expected recoils for a given experiment
and WIMP mass, we integrate Eqn.~(\ref{eqn:dRdE}) over the nucleus
recoil energy to find the recoil rate $R$ per unit detector mass:
\begin{subequations} \label{eqn:rate}
\begin{equation} \label{eqn:rateone}
 R(t) = \int_{E_{1}/Q}^{E_{2}/Q} dE \,
        \epsilon(E) \frac{\rho}{2 m \mu^2} \, \sigma(q) \, \eta(E,t) .
\end{equation}
$\epsilon(E)$ is the (energy dependent) efficiency of the experiment,
due, e.g., to data cuts designed to reduce backgrounds.  $Q$ is the
quenching factor relating the observed energy $E_{det}$ (in some cases
referred to as the electron-equivalent energy) with the actual recoil
energy $E_{rec}$: $E_{det} = Q E_{rec}$.  The energy range between
$E_{1}$ and $E_{2}$ is that of \textit{observed} energies for some data
bin of the detector (where experiments often bin observed recoils by
energy).
Note some single element detectors can calibrate their energy scales
to $E_{det} = E_{rec}$, in which case the quenching factor can be
ignored.  For detectors with multiple elements, the total rate is
given by:
\begin{equation} \label{eqn:ratetot}
 R_{tot}(t) = \sum_i f_i R_i(t)
\end{equation}
\end{subequations}
where $f_i$ is the mass fraction and $R_i$ is the rate
(Eqn.~(\ref{eqn:rateone})) for element $i$.

The expected number of recoils observed by a detector is given by:
\begin{equation} \label{eqn:recoils}
 N_{rec} = M_{det} T R
\end{equation}
where $M_{det}$ is the detector mass and $T$ is the exposure time.
To calculate the constants in Eqn.~(\ref{eqn:conic}), we first
rewrite Eqn.~(\ref{eqn:SDCS}) as:
\begin{equation} \label{eqn:SDCS2}
 \sigma_{SD}(q) = h_{pp}(q) a_p^2 + h_{pn}(q) a_p a_n
               + h_{nn}(q) a_n^2
\end{equation}
where
\begin{equation} \label{eqn:SDCSij}
 h_{ij}(q) \equiv \frac{32 \mu^2 G_F^2}{2 J + 1} S_{ij}(q)
\end{equation}
is independent of the couplings $a_p$ and $a_n$.  The constants
$A_{rec}$, $B_{rec}$, and $C_{rec}$ in Eqn.~(\ref{eqn:conicrec}) are
then determined from Eqn.~(\ref{eqn:recoils}) by replacing $\sigma(q)$
in Eqn.~(\ref{eqn:rateone}) with the appropriate $h_{ij}(q)$, e.g.\ 
\begin{equation} \label{eqn:Arec}
 A_{rec} = M_{det} T
           \int_{E_{1}/Q}^{E_{2}/Q} dE \,
                \epsilon(E) \frac{\rho}{2 m \mu^2} \, h_{pp}(q)
                \, \eta(E,t) .
\end{equation}
Similarly, $B_{rec}$ is given by Eq.(\ref{eqn:Arec}) but with
$h_{pp}$ replaced by $h_{pn}$, and $C_{rec}$ is given by
Eq.(\ref{eqn:Arec}) but with $h_{pp}$ replaced by $h_{nn}$.

For large detectors that can obtain high statistics, the annual
modulation of the rate (due to the rotation of the Earth about the Sun)
may be detectable, where
\begin{equation} \label{eqn:modulation}
 R(t) \approx R_0 + R_m \cos(\omega t) .
\end{equation}
The rate (for most energies and for a Maxwellian velocity distribution)
is maximized around June 2 and minimized around December 2; this phase
is reversed at high enough energies \cite{Lewis:2003bv}.  The
modulation amplitude,
\begin{equation} \label{eqn:modampl}
 N_{ma} \equiv R_m = \frac{1}{2} \left[ R(\textrm{June 2})
                                        - R(\textrm{Dec 2}) \right],
\end{equation}
denoted by $N_{ma}$ for consistency with Eqn.~(\ref{eqn:conicrec}),
can be put in the form of Eqn.~(\ref{eqn:conicmod}) in the same manner
as the total recoils above.  Hence $A_{ma}$ is the value
of Eqn.~(\ref{eqn:modampl}) when $\sigma(q) \to h_{pp}(q)$ in
Eqn.~(\ref{eqn:rateone}):
\begin{eqnarray} \label{eqn:Ama}
 A_{ma} &=& \int_{E_{1}/Q}^{E_{2}/Q} dE \, \epsilon(E)
                 \frac{\rho}{2 m \mu^2} \, h_{pp}(q) \nonumber\\
        & & \qquad \times \, \frac{1}{2} [\eta(E,\textrm{June 2})
                                          - \eta(E,\textrm{Dec 2})] .
\end{eqnarray}
Similarly, $B_{ma}$ is given by Eq.(\ref{eqn:Ama}) but with
$h_{pp}$ replaced by $h_{pn}$, and $C_{ma}$ is given by
Eq.(\ref{eqn:Ama}) but with $h_{pp}$ replaced by $h_{nn}$.


\end{document}